\RequirePackage{fix-cm}
\documentclass[smallextended]{svjour3}       
\smartqed  
\usepackage{graphicx}
\usepackage{epstopdf}
\usepackage{tikz}
\usepackage{float}
\usepackage{float}
\usepackage{hyperref}
\usepackage{soul}
\usepackage{amsmath}

\newcommand{\ket}[1]{$|#1\rangle$}

\begin{document}

\title{Landau-Zener Transitions in Spin Qubit Encoded in Three Quantum Dots}


\author{Jakub {\L}uczak \and Bogdan R. Bu{\l}ka
}


\institute{Institute of Molecular Physics, Polish Academy of Sciences, M. Smoluchowskiego 17, 60-179
Pozna{\'n}, Poland\\ \email{jakub.luczak@ifmpan.poznan.pl}}

\date{Received: date / Accepted: date}

\maketitle

\begin{abstract}
We study generation and dynamics of an exchange spin qubit encoded in three coherently coupled quantum dots with three electrons. For two geometries of the system a
linear and a triangular one, the creation and coherent control of the qubit states are performed by the Landau--Zener transitions. In the triangular case both the qubit states
are equivalent and can be easily generated for particular symmetries of the system.
If one of the dots is smaller than the others one can observe Rabi oscillations, that can be used for coherent manipulation of the qubit states. The linear system is easier to fabricate; however, then the qubit states are not equivalent, making qubit operations more difficult to control.
\end{abstract}

\section{Introduction}

Recent progress in the experimental realization of the semiconductor quantum dots (QDs) gives the tool to perform compatible and fully scalable systems
needed
in the quantum computations. In contrast to charge qubits, spin qubits are characterized by long decoherence times necessary in the quantum
computation \cite{vrijen}. To encode the qubit in the single electron spin in QD one needs to apply  a magnetic field which removes the spin degeneracy. Control of
the spin
qubit can be performed by the electron spin resonance (ESR) \cite{engel,nowack}. The read-out of the final qubit state can be done using transport measurements in the Pauli spin blockade regime with an auxiliary QD or with a quantum point contact (QPC) \cite{veldhorst}.

There are proposals \cite{loss,petta2005,bluhm,sarkka,granger,dial,wu,barthel,maune} to build the qubit in a two spin system in double quantum dots (2QD). The qubit
logical subspace is defined by a singlet (S) and one of triplets ($T^{S_Z}$), which
correspond to the north and the south pole of the Bloch sphere, respectively. Applying an external magnetic field one removes degeneration between the triplets
and the information is stored in the $S-T^{+ 1}$ subspace. The preparation and manipulation of the qubit can be done by fast electrical pulses (in a nanosecond scale) which change an exchange
interaction between the spins \cite{petta2005}. The control of the qubit is performed by the Landau--Zener (L--Z)
transition \cite{shevchenko} through an anticrossing point in a non-adiabatic regime. The anticrossing comes from mixing between the singlet and triplet states due to nuclear hyperfine fields
\cite{petta2005,bluhm,sarkka}, a spin-orbit coupling \cite{hanson}, or an
inhomogeneous magnetic field \cite{wu}. The mixing is needed for proper functioning of the $S-T^{+1}$ qubit, however it can cause some unwanted decoherence
processes.
The L--Z method can be used to implement the universal quantum gates
with high fidelity \cite{bason,hicke} and to measure the $S-T^{+1}$ splitting when the spin-orbit coupling and hyperfine interaction compete with each other \cite{nichol}.
The 2QD system with two spins allows also to encode the qubit in the $S-T^{0}$ subspace \cite{petta2005,dial,wu,barthel}. In the external magnetic field
the qubit state $T^{0}$ is the excited state, therefore one needs to pass through the $S-T^{+1}$ anticrossing quick enough to remain in the qubit state $S$. The mixing of $S$
and $T^{0}$ states is induced in an inhomogeneous magnetic field and the initialization can be performed with high fidelity \cite{barthel}. Moreover both the
qubit states have $S_Z=0$, therefore they are unaffected by noises in an uniform magnetic fields.

DiVincenzo {\it et al.} \cite{vincenzo} proposed an exchange-only qubit in three spin system in a triple quantum dot (TQD) device.
An advantage of the proposal is encoding the qubit in the doublet states with the same spin z-component ($S_Z$).
It was pointed out \cite{dfs} that the doublet subspace is immune to the decoherence processes. In the system, the
full unitary operations of the qubit states are done by purely electrical control of the exchange interactions between the spins.
Recently, TQD in a linear geometry has been investigated, both theoretically \cite{fei,pal,busl} and experimentally \cite{busl2013,aers,laird10,gaudreau}.
Another proposition is a resonant exchange qubit \cite{taylor,medford}, where the manipulation is done by applying an rf gate-voltage pulses to one of the
gates. If the oscillation frequency is matched to the exchange interaction one can observe the nutations between the qubit states.
The DiVincenzo scheme is not
limited only to the linear TQD system. Shi {\it et al.} \cite{shi} proposed an electrically controlled hybrid qubit encoded in 2QD with many levels and three spins.

Recent theoretical studies \cite{hawrylak,bulka,luczak} showed advantages encoding of the qubit on TQD with a triangular geometry. In this case
both the doublet states are equivalent and can be easily controlled by changing the TQD symmetry. Single qubit operations, the read-out and the decoherence related with an
external electrodes as well as leakage processes were studied as well \cite{luczak}. The triangular TQD devices were fabricated in the three
lateral quantum dots by the atomic force microscope \cite{rogge}, the electron-beam lithography
\cite{seo} and in the vertical quantum dots \cite{amaha09}. Similar structures can be found in molecular magnets \cite{molmag}
which exhibit rich quantum dynamics.

In this paper we would like to show how one can encode the spin qubit and study its dynamics by means of the L--Z transitions  for different symmetries of the
TQD system. We model the system within the Hubbard Hamiltonian where the symmetry is fully electrically controlled by the local gate potentials applied to the
quantum dots.
Two geometries of TQD will be taken into consideration, the
linear and the triangular one, for which one expects significant differences in qubit generation and its dynamics. The linear case is related to the
experimental
papers \cite{aers,laird10} where a qubit state was initialized in the doublet subspace by an adiabatic passage. They observed a coherent rotation between the qubit states when  an exchange pulse applied to the system induced the L--Z transition.
We would like to extend the investigation on
dynamic generation of the qubit states and study conditions for qubit encoding. The main purpose is to study the triangular TQD where one expects that both the
qubit states can be easily generated by the L--Z transition. We will examine different symmetries of the system to generate any qubit state on the Bloch sphere.
Next, the evolution of the qubit states with time dependent gate potentials will be analyzed. We expect that the qubit states could be
degenerated for a special condition (when a pseudo-magnetic field vanishes). The L--Z passage through this point could lead to Rabi oscillations which can be used for
coherent qubit manipulations. We will show that this effect can be observed for some special symmetry: the triangular TQD with one of the dots being smaller.

\section{Model of the system}

We investigate an artificial molecule built on three coherently coupled quantum dots (TQD) which is described by the Hubbard Hamiltonian
\begin{eqnarray}\label{modelHubbardEPt}
  \hat{H} &= \sum_{i,\sigma}\tilde{\epsilon}_{i}(t)\,n_{i\sigma}+\sum_{i<j,\sigma}t_{ij}(c^{\dagger}_{i\sigma}c_{j\sigma}
  +h.c.)  \\ \nonumber &+\sum_{i}U_in_{i\uparrow}n_{i\downarrow}- g \mu_B B_Z \sum_i S_{Z,i}.
\end{eqnarray}
Here, $\tilde{\epsilon}_{i}(t)$ is a time dependent local site energy controlled by a gate potential applied to the $i$-th quantum dot, $t_{ij}$ is a
hopping parameter between the dots, $U_i$ describes a intra-dot Coulomb interaction. We assume that TQD is placed in an external magnetic field $B_Z$ which splits the
spin levels due to the Zeeman effect, $g$ is the Land\'{e} g-factor for an electron and $\mu_B$ is the magnetic moment.

Confining to three electrons in the TQD system one can construct two subspaces of the states: quadruplets and doublets. The quadruplets  are the states with the
total spin
$S=3/2$ and have the form
\begin{eqnarray}\label{Q32}
|Q^{+3/2}\rangle&=&|\uparrow_1\uparrow_2\uparrow_3\rangle,\\ \label{Q12}
|Q^{+1/2}\rangle&=&\frac{1}{\sqrt{3}}(|\uparrow_1\uparrow_2\downarrow_3\rangle
+|\uparrow_1\downarrow_2\uparrow_3\rangle+|\downarrow_1\uparrow_2\uparrow_3\rangle),
\end{eqnarray}
for $S_Z=\{+3/2, +1/2\}$, and similarly for the states with $S_Z=\{-3/2, -1/2\}$ (flipping all  spins in   (\ref{Q32})-(\ref{Q12})).

We focus on the second subspace formed by the doublets with $S=1/2$ and $S_z=\pm1/2$. For $S_z=+1/2$ the doublets can be expressed as:
\begin{eqnarray}\label{psi}
|\Psi^{+1/2}_D(t)\rangle&=\alpha^{D_S}(t)|D_S^{+1/2}\rangle+\alpha^{D_T}(t)|D_T^{+1/2}\rangle\\ \nonumber &+\sum_{i\neq j}\alpha^{iij}(t)
|D^{+1/2}_{iij}\rangle,
\end{eqnarray}
where
\begin{eqnarray}\label{ds}
|D_S^{+1/2}\rangle&=\frac{1}{\sqrt{2}}(|\uparrow_1\uparrow_2\downarrow_3\rangle-|\downarrow_1\uparrow_2\uparrow_3\rangle) \equiv
|\uparrow_2\rangle\otimes|S_{13}\rangle, \\
\label{dt}
|D_T^{+1/2}\rangle
&=\frac{1}{\sqrt{6}}(|\uparrow_1\uparrow_2\downarrow_3\rangle+|\downarrow_1\uparrow_2\uparrow_3\rangle -2|\uparrow_1\downarrow_2\uparrow_3\rangle)\nonumber\\
&\equiv\frac{1}{\sqrt{3}}|\uparrow_2\rangle\otimes|T_{13}^0\rangle-\sqrt{\frac{2}{3}}|\downarrow_2\rangle\otimes |T_{13}^{+1} \rangle,
\end{eqnarray}
are the states constructed by adding an electron to the dot 2 when two other electrons formed the singlet
$|S_{13}\rangle=(|\uparrow_1\downarrow_3\rangle-|\downarrow_1\uparrow_3\rangle)/\sqrt{2}$ and the triplets
$|T_{13}^{0}\rangle=(|\uparrow_1\downarrow_3\rangle+|\downarrow_1\uparrow_3\rangle)/\sqrt{2}$, $|T_{13}^{+1}\rangle=|\uparrow_1\uparrow_3\rangle$ on the bond 13, respectively.
The states $|D_S^{+1/2}\rangle$ and $|D_T^{+1/2}\rangle$ are used to encode the qubit according to DiVincenzo {\it et al.} scheme \cite{vincenzo}.
The last term in (\ref{psi}), with  $|D^{+1/2}_{iij}\rangle=|\uparrow_i\downarrow_i\uparrow_j\rangle$, corresponds to the state  with double occupancy of one of the dot and will be used to the qubit initialization. To simplify the notation we omit the spin index $+1/2$
in the future consideration.

The model (\ref{modelHubbardEPt}) neglects hyperfine interactions and a spin-orbit coupling, which means that the doublet and quadruplet subspaces are separated. Therefore the current studies of qubit dynamics can be confined to the doublet subspace only. The model can be applied to the Si-based quantum dots, the systems which are very promising in quantum computation due to long decoherence and relaxation times which can be in the order of a few seconds \cite{zwanenburg,simmons}.

\subsection{Qubit encoded in the doublet subspace}

Let us focus on the spin qubit states \ket{D_S} and \ket{D_T} which can be presented on the Bloch sphere as the north and south pole, respectively. In the
Hubbard model the exchange interactions are generated by virtual transitions to the excited double occupied states. To see these processes we perform the
perturbative canonical transformation \cite{kostyrko} of the Hubbard Hamiltonian (\ref{modelHubbardEPt}) to an effective Heisenberg Hamiltonian
\begin{eqnarray}\label{modelHeisenberga}
\hat{H}_{eff} &=\sum_{i<j}J_{ij}
(\mathbf{S}_i\cdot\mathbf{S}_j- \frac{1}{4})- g \mu_B B_z \sum_i S_{z,i},
\end{eqnarray}
where the superexchange couplings $J_{ij}$ are derived out to the second order
treating the interdot hoppings as small ($t_{ij}\ll U$)
\begin{eqnarray}\label{heisenbergJ}
J_{ij}&=2|t_{ij}|^2\left(\dfrac{1}{U_j + \tilde{\epsilon}_j - \tilde{\epsilon}_i}+\dfrac{1}{U_i + \tilde{\epsilon}_i - \tilde{\epsilon}_j}\right).
\end{eqnarray}
Notice that this Hamiltonian works only for the doublets with single occupation of the quantum
dots.

The Hamiltonian of the spin qubit encoded in the doublet subspace \ket{D_S} and \ket{D_T} can be
expressed as
\begin{eqnarray}\label{hamilpaulixyz}
\hat{H_\sigma}=-\frac{1}{2}(3J+g \mu_B B_z)\mathbf{1}+\frac{\delta}{2}\sigma_z+\frac{\gamma}{2}\sigma_x,
\end{eqnarray}
where \begin{eqnarray}\label{jjj}
J &= \frac{1}{3}(J_{12}+J_{23}+J_{31}), \\ \label{delta}
\delta &= \frac{1}{2}(J_{12}+J_{23}-2J_{31}),\\ \label{gamma}
\gamma &= \frac{\sqrt{3}}{2}(J_{12}-J_{23}).
\end{eqnarray}
Note that $\delta$ and $\gamma$ can be interpreted as the z and x component of a pseudo- magnetic field ${\bf b}=(\gamma,0,\delta)$. Eigenvalues of the
Hamiltonian (\ref{hamilpaulixyz}) are
\begin{eqnarray}\label{wartwlasne}
E_{D}^\pm= -\frac{3}{2}J-\frac{g \mu_B B_z}{2} \pm \frac{\Delta}{2},
\end{eqnarray}
where $J$ is the energy separation of the quadruplet and the doublet subspaces, and $\Delta=\sqrt{\gamma^2+\delta^2}$ is the energy gap between the doublets.
For a system of Si/SiGe quantum dots, the gap can be of the order of $\Delta \sim 21.6$ $\mu$eV \cite{shi2013}. The corresponding eigenstates are
\begin{eqnarray}\label{psi_m}
|D^-\rangle= \cos \phi |D_S\rangle +\sin \phi |D_T\rangle ,\\ \label{psi_p}
|D^+\rangle= \sin \phi |D_S\rangle-\cos \phi |D_T\rangle ,
\end{eqnarray}
where $\phi= \arctan (\gamma/(\delta-\Delta))$ denotes the angle of the pseudo-field ${\bf b}$ with respect to the z-axis.

In this paper we consider two configurations of the TQD: the triangular one (see fig. \ref{modelTQDTrig}), with  all superexchange interactions are always on
($J_{ij}\neq 0$), and the linear one for which the outermost spins are decoupled (we take $J_{31}=0$ - see linear TQD scheme in fig. \ref{LZS_lin}a).
From  equation (\ref{delta}) one can see that for the linear molecule $\delta = (J_{12}+J_{32})/2 >0$, therefore the state \ket{D_T} is energetically favorable.
For the symmetric case with $J_{12}=J_{23}$ the mixing $\gamma = (3/2)(J_{12}-J_{23})=0$ and the field $\bf{b}$ is oriented toward the south pole, with
\ket{D_T}
as the ground state.
For the triangular TQD all exchange couplings are equivalent. Therefore, by proper manipulation of the
exchange couplings one can change the triangular symmetry and get any orientation of the pseudo-field ${\bf b}$ on the x-z plane.
This effect
is used to prepare the qubit state on the Bloch sphere in any superposition given by $|D^{\pm}\rangle$.

\section{L--Z effect in the linear molecule}

\begin{figure}
\centering
\includegraphics[width=.65\textwidth,clip]{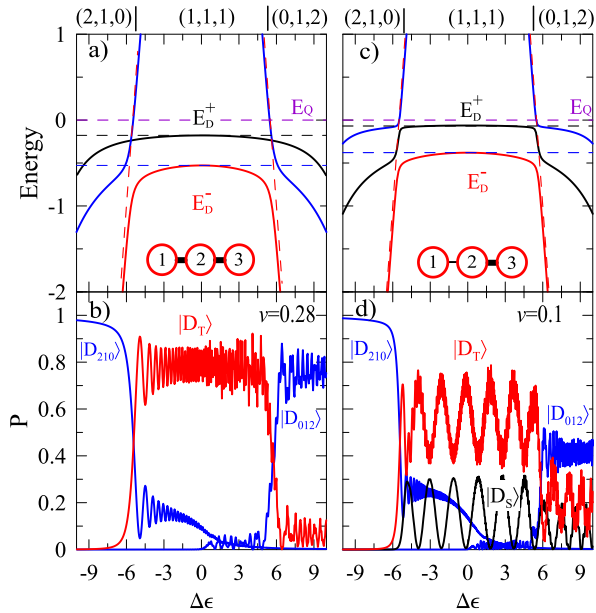}
\caption{Landau--Zener transitions for the linear TQD molecule with symmetric couplings $t_{12}=t_{23}=-1$ (left panel) and asymmetric
couplings
$2t_{12}=t_{23}=-1$ (right panel),  for $U_i=11$. Figure a) and c) present the adiabatic (solid lines) and diabatic (dashed lines) evolution of energy levels with respect to
the detuning parameter
$\Delta\epsilon=\tilde{\epsilon}_1-\tilde{\epsilon}_3$. The solid red curve corresponds to the ground state energy, the solid black curve presents the
first excited doublet state, while the dashed violet line is for the
quadruplet which is independent of the detuning parameter. The charge states $(N_1,N_2,N_3)$ of the system are marked above the plots. Notice that the spin qubit operates in the charge configuration (1,1,1) and is encoded in the states \ket{D_S} and \ket{D_T} with the fidelity $F=0.95$. The lower panels, figure
b) and d), present the occupation probabilities of various states as a time dependent function (with respect to $\Delta\epsilon=2v t$) for the rate $v=0.28$ (b)
and  $v=0.1$ (d). For better clarity we plotted only states with main contribution. Notice that in the symmetric case \ket{D_S} is not
occupied. For the initial time $\Delta\epsilon(t_{in})=-10$ the system was in the ground state \ket{D_{210}}.}
\label{LZS_lin}
\end{figure}

First we consider the dynamics in the linear TQD (see the scheme in fig. \ref{LZS_lin}a) which was studied recently by
several experimental groups \cite{aers,laird10,gaudreau}. The paper \cite{laird10} showed the initialization of the qubit state by an adiabatic passage between
different charge regions. However there was not presented how the qubit state was encoded on the Bloch sphere.
We want to expand these studies and find conditions for the generation of the qubit states \ket{D_S} and \ket{D_T} by means of the L--Z transition.
The evolution of the doublet state $|\Psi_D(t)\rangle$ (\ref{psi}) will be considered for different symmetries of the system for which one can expect
initialization of a various qubit states on the Bloch sphere.
Let us stress that our model (\ref{modelHubbardEPt}) conserves the total spin, the system remains in the doublet subspace during the evolution. The advantage of this approach is to operate in the  decoherence-free subspace (DFS) \cite{dfs}.

In analogy to the experiments \cite{aers,laird10,gaudreau} we assume that applying time dependent local gate potentials the site energies
$\tilde{\epsilon}_i$ of the
quantum dots are changed as
\begin{eqnarray}
\tilde{\epsilon}_1(t)&=&\epsilon_1 + v\,t, \\ \nonumber
\tilde{\epsilon}_2(t)&=&\epsilon_2, \\ \nonumber
\tilde{\epsilon}_3(t)&=&\epsilon_3 - v\,t,
\end{eqnarray}
where $v$ is a speed rate of the potential changes. Taking $\epsilon_i=0$ the detuning between the levels in the outermost quantum dots is given  by
$\Delta\epsilon=\tilde{\epsilon}_1(t)-\tilde{\epsilon}_3(t)= 2 v\,t$. This allows to change the number of electrons $N_i$ in the quantum dots and
transfer between different charge configurations $(N_1,N_2,N_3)$ with the total number of electrons $\sum_{i}N_i=3$.
The dynamics of the system is described by the time dependent Schr\"{o}dinger equation
\begin{eqnarray}\label{schrodinger}
i \frac{d}{dt}|\Psi_D(t)\rangle=\hat{H}|\Psi_D(t)\rangle,
\end{eqnarray}
with the Hubbard Hamiltonian (\ref{modelHubbardEPt}) and the doublet base given by (\ref{psi}). The equation (\ref{schrodinger}) neglects decoherence and relaxation processes which can be caused by e.g. a charge noise or spin-flip precesses. The charge noises are related with fluctuations of a local electrical potential in the quantum dot confinement and tunnel barriers. The corresponding decoherence time can be of the order of microseconds \cite{culcer}. In the system connected to external electrodes one can observe charge fluctuations as well as spin--flip processes (with the relaxation time around 100 ns \cite{luczak}). We show later that the equation (\ref{schrodinger}) very well describes dynamics of the system if the speed rate is taken reasonable large and the time of the qubit generation is much shorter than the relaxation and decoherence times.

We are interested in the L--Z transition  between the charge configurations (2,1,0) to (1,1,1) when the qubit states \ket{D_S} and \ket{D_T} are generated. Since the calculations are performed within the Hubbard model the qubit states are not perfect, they are always in a small superposition with the double occupation states.
The accuracy of the initialization operation of the qubit can be defined by a fidelity,  which is a measure of closeness of two quantum states and is given by $F=Tr \left[\sqrt{\rho_i} \rho_r \sqrt{\rho_i}\right]$, where $\rho_i$ is a density matrix of a desired ideal state and $\rho_r=|\Psi_D\rangle \langle\Psi_D|$ is the real final state \cite{hauke}. For a perfect qubit generation the fidelity is unity, however due to non-qubit states the fidelity is lowered.  For large $U_i$ the separation between the double occupied states and the qubit states is larger, and the fidelity goes to unity, however the exchange interactions $J_{ij}$ decrease and the gap between the doublet states becomes smaller.

The results of the numerical calculations of the evolution of the doublet states and the energy levels as a function of the detuning parameter are presented in Figure \ref{LZS_lin} for the linear TQD with the symmetric tunnelings between the dots
$t_{12}=t_{23}=-1$ (the left panel) and for the asymmetric case with
$2t_{12}=t_{23}=-1$ (the right panel).
Fig.\ref{LZS_lin}a) presents the adiabatic and diabatic energy levels (full and dashed curves, respectively) versus the detuning $\Delta\epsilon$ for the
symmetric case. The solid red curve describes the evolution of the ground state, from the state \ket{D_{210}} via \ket{D_T} (in the (1,1,1) region) to the final
state \ket{D_{012}}.
The dashed curves correspond to the diabatic energy levels: the blue and the red curve for \ket{D_T} and \ket{D_{210}}, respectively, while
the black dashed curve
describes the diabatic energy level \ket{D_S}.
Notice that for the symmetric coupling between the dots the ground state in the region (1,1,1) is \ket{D_T} and the second doublet \ket{D_S} can not be
generated. Optimal generation of the qubit is for the adiabatic passage for which the fidelity $F\approx0.95$ at $\Delta \epsilon=0$. The fidelity can be higher for larger $U_i$ when the occupation of the non-qubit states is negligible.

For the case presented in Fig.~\ref{LZS_lin} we have the L-Z transitions in a many level system. The probability that the system remains at the same state can be estimated as \cite{kayanuma}
\begin{eqnarray}\label{plz}
P_{LZ}=\prod_{k=1}^N p_{0k}=\exp\left(-2\pi \sum_{k=1}^N \frac{\Delta E_{0k}^2}{4v}\right).
\end{eqnarray}
The result is derived for $N$ final states and under the condition that the system stays in the state \ket{0} at the final time $t_{fin}=+\infty$ if it has started from \ket{0} at
$t_{in}=-\infty$.
The probability $p_{0k}$ to remain in \ket{0} after crossing with the state \ket{k} is regarded as an independent process.
Here we denote $\Delta E_{0k}$ as the gap at the anticrossing point with the state \ket{k}, and $\hbar=1$. Similarly one can find the probability to generate
the system in the  first state as $P^{gen}_{|1\rangle}=1-p_{01}$ and in the $n$-th state ($n\geq2$) as $P^{gen}_{|n\rangle}=(1-p_{0n})\prod_{k=1}^{n-1}p_{0k}$.

The numerical calculations of the time evolution of the occupation probabilities of the doublet states are showed in Fig.\ref{LZS_lin}b). At an initial
moment $\Delta\epsilon(t_{in})=-10$ the system was in the ground state \ket{D_{210}}.
The calculations was performed for the speed rate $v=0.28$. Notice that this speed rate is not optimal for the qubit generation and is taken to show the dynamics in the (1,1,1) region clearly. After the crossing point (for $\Delta\epsilon> -U/2$) one can observe the L--Z transition to the
qubit state \ket{D_T} with characteristic oscillations in the occupation probabilities for
\ket{D_T} and \ket{D_{210}}. The period of the oscillations is inversely proportional to the energy gap. At
the transition point the energy gap is $\Delta E_{01}\approx0.66$ and hence one can estimate the generation probability $P^{gen}_{D_T}\approx 0.70$.
Notice that the second qubit state \ket{D_S} is not generated.
When the potential becomes larger ($\Delta\epsilon >U/2$) the system goes through the other L--Z passage, to
the charge configuration (0,1,2).

Let us estimate the sweep time $t^{sw}$ needed to transfer the system from the initial state to the center of the (1,1,1) region. The hopping parameter
determined in the experiment \cite{gaudreau} is $t_{ij}\approx 10$ $\mu$eV. The energy gap can be estimated as $\Delta E_{01}= 6.6$ $\mu$eV and the speed rate in our calculations is
$v\approx 43$ keV/s. Hence the qubit is generated in the time $t^{sw}\approx 2.31$ ns with the probability $P_{|D_T\rangle}^{gen}\approx0.70$. The estimated time is much shorter than spin-flip relaxation \cite{luczak} and decoherence caused by the charge noise \cite{culcer}, which means that our approximation (\ref{schrodinger}) is justified.

For the asymmetric couplings between the quantum dots, which mimics the experimental situation \cite{laird10}, one can get observe some mixing of the qubit state \ket{D_T} with \ket{D_S} - see the right panel in Fig.~\ref{LZS_lin}. In the
considered case we take $2t_{12}=t_{23}=-1$ and the superexchange couplings are asymmetric $4J_{12}=J_{23}=4t^2_{23}U/(U^2-v^2t^2)$. The qubit parameters are
$\delta=5J_{23}/8$ and $\gamma=-3\sqrt{3}J_{23}/8$, therefore the  L--Z transition can generate two states \ket{D^-} and \ket{D^+} which are superpositions of
\ket{D_S} and \ket{D_T}.
One can see a double anticrossing in the adiabatic energy levels in Fig.\ref{LZS_lin}c between the blue-red and the blue-black curves, which
correspond the L--Z transitions: $|D_{210}\rangle\to |D^-\rangle$ and $|D_{210}\rangle\to |D^+\rangle$. The energy gaps at these points are $\Delta
E_{01}\approx
0.4$ and $\Delta E_{02}\approx0.55$, respectively. For the speed rate $v=0.1$, which provides the same $\Delta E_{01}^2/v$ rate like in the previous case, one
can generate these states with the probabilities $P^{gen}_{|D^{-}\rangle}\approx 0.70$ and $P^{gen}_{|D^{+}\rangle}\approx 0.259$. The transition time can
be estimated as $t^{sw}\approx 6.3$ ns. The generated qubit rotates on
the Bloch sphere what is seen in fig. \ref{LZS_lin}d) as strong oscillations between the states \ket{D_S} and \ket{D_T}.

\begin{figure}
\centering
\includegraphics[width=.45\textwidth,clip]{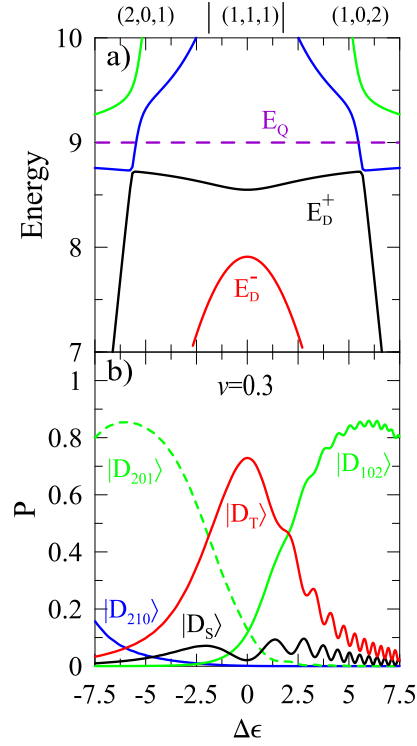}
\caption{Landau--Zener transitions for the linear molecule with a small central dot for the symmetric tunneling $t_{12}=t_{23}=-1$. The adiabatic evolution of
the energy levels is plotted in fig. a) as solid curves, the purple-dashed curve corresponds to the energy of quadruplets states. Below  fig. b) presents the
time evolution of the probability of the doublet states with the speed rate $v=0.3$. The other parameters are: $\epsilon_2=9$, $\epsilon_{1,3}=0$, $U_{1,3}=11$
and $U_2=24$.}
\label{LZS_lin2}
\end{figure}

Let us consider the TQD system with the central dot smaller than two others, and the L--Z transitions between the charge configurations (1,0,2)$\leftrightarrow$(1,1,1)
and (2,0,1)$\leftrightarrow$(1,1,1). The similar passages were studies in the experiments  \cite{aers,laird10,gaudreau}. For the model (\ref{modelHubbardEPt}) we
take $\epsilon_2>\epsilon_1=\epsilon_3=0$ and $U_2>U_1=U_3=U$. For this case the symmetry is broken, and from eq. (\ref{heisenbergJ}) one gets asymmetric
superexchange couplings $J_{12}\neq J_{23}$ for any $\Delta \epsilon\neq 0$, which leads to mixing between the qubit states \ket{D_T} and \ket{D_S}.
Figure \ref{LZS_lin2} presents dynamics of the system for $\epsilon_2=9$, $U_2=24$ and the symmetric electron hopping between the dots. At an initial moment
$\Delta\epsilon=-7.5$ and the ground state is \ket{D_{201}} with a superposition of \ket{D_{210}}.
Figure \ref{LZS_lin2}a), shows the adiabatic evolution of the energy levels which is qualitatively similar as observed in the experiment \cite{aers}. For a given speed rate $v=0.3$ the L--Z transition generates the state \ket{D^-}, whereas the state \ket{D^+} is not generated - see Fig.\ref{LZS_lin2}b). Notice that at $\Delta\epsilon=0$, there is no
mixing
between the qubit states ($\gamma=0$), therefore the occupation probability of \ket{D_S} decreases and goes to 0 for the adiabatic passage. Moreover, one can
see
that the qubit state \ket{D^-} is  the superposition with the states \ket{D_{201}} and \ket{D_{102}}, which also leads to a deep in the adiabatic curve $E_D^+$
(the
black one in Fig. \ref{LZS_lin2}a)). The fidelity $F\approx 0.74$ is rather low, which means that the qubit can not be so good generated as in the previous cases.

\section{L--Z effect in the triangular molecule}

\begin{figure}[ht]\centering
\includegraphics[width=0.55\textwidth,clip]{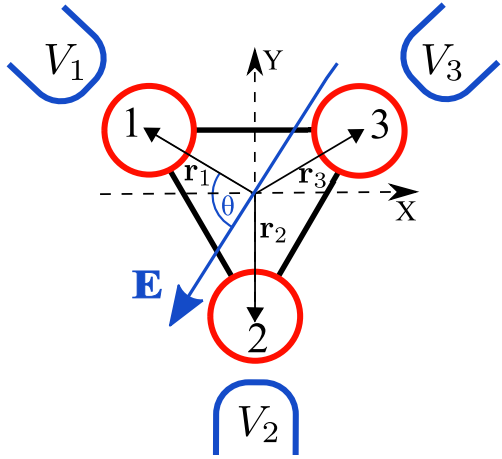}
\caption{Scheme of the TQD in the triangular geometry in the presence of effective electric field caused by the potentials $V_i$ applied to the gate electrodes.}
\label{modelTQDTrig}
\end{figure}

Let us now consider the L--Z transitions in the TQD with the triangular geometry (shown
in figure \ref{modelTQDTrig}) where the local gate electrodes control the position of the energy levels, $\tilde{\epsilon}_i=\epsilon_i+eV_i$ at each dot.
Here we are interested in an influence of the triangular symmetry on the energy structure of the qubit and its dynamics.
To describe the symmetry breaking effect in TQD, it is more convenient to introduce an effective in-plain electric field ${\bf E}$, instead to use the local
gate potentials $V_i$ as the parameters. For a small value of {\bf E} the energy level of the $i$-th quantum dot can be expressed as
\begin{eqnarray}
\tilde{\epsilon}_i=\epsilon_i+g_E \cos\left[\theta-\left(i-1\right)\frac{2\pi}{3}\right],
\end{eqnarray}
where $g_E= e|{\bf E}||{\bf r}_1+{\bf r}_2|/2$ is the parameter describing electron polarization, $e$ is the elementary electron charge, ${\bf r}_i$ - the
vector
showing the position of the $i$-th quantum dot in the coordinate system, $\theta$ is the angle between ${\bf E}$
and the vector ${\bf r_1}$. We would like to study the L--Z passages generated by a time dependent electric field described by the parameter
$g_E\equiv v t$ at a fixed angle $\theta$.

\begin{figure}
\centering
\includegraphics[width=.9\textwidth,clip]{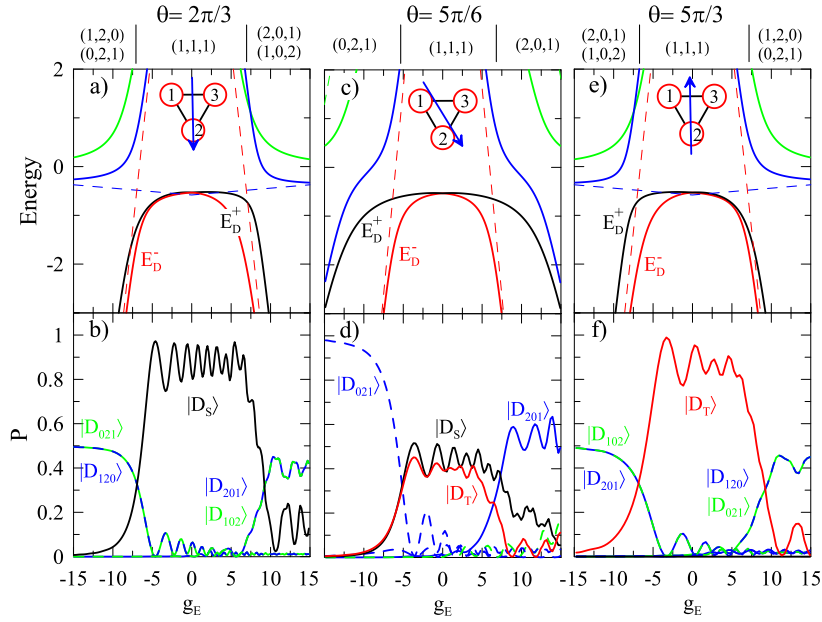}
\caption{Landau--Zener transitions  in TQD with the triangular geometry for three orientations of the electric field: $\theta=2\pi/3$ (left panel),
$\theta=5\pi/6$
(middle panel) and $\theta=5\pi/3$ (right panel) - see inserts. The energy spectrum is presented in the top panels where the dashed and the solid curves correspond
to the diabatic and the adiabatic passages, respectively. In the adiabatic regime the fidelity $F\approx 0.95$. Bottom panels show the time evolution of the occupation probabilities of the doublet states, where $g_E = v t$. In
calculations we take: $U=11$, $t_{ij}=-1$, $\epsilon_{i}=0$.  Plots are taken for the same $\Delta E_{01}^2/v$ rate as in fig.~\ref{LZS_lin}: b)
$\Delta E_{01} \approx 2.3$ and $v=2.2$; d) $\Delta E_{01} \approx 2.28$ and $v=3.48$; f) $\Delta E_{01} \approx 3.47$ and $v=5.01$.}
\label{LZS_trig}
\end{figure}

Figure \ref{LZS_trig} shows dynamics in TQD with the triangular geometry for three different orientations of the electric field:
$\theta=2\pi/3$,  $\theta=5\pi/6$ and $\theta=5\pi/3$ (in the left, middle and right panel, respectively).
For $\theta=2\pi/3$ the electric field is toward the dot 2, and at the initial time, the electron polarization parameter is $g_E(t_{in})=-15$ for which the dot
levels are
$\tilde{\epsilon}_2+U_2<\tilde{\epsilon}_1=\tilde{\epsilon}_3$.
The ground state $|\psi_{in}\rangle=(|D_{021}\rangle+|D_{120}\rangle)/\sqrt{2}$ is a superposition of the charge configuration (0,2,1) and (1,2,0)
(with an equal charge redistribution between the dot 1 and 3).
In Fig. \ref{LZS_trig}a) one can see the anticrossing point between the adiabatic energy levels (the red and
blue solid curves) corresponding to the L--Z transition to the region (1,1,1) with the ground state $|D^-\rangle=|D_S\rangle$. The energy gap at the
anticrossing
point is $\Delta E_{01}\approx2.3$ and for the speed rate $v=2.2$  the state \ket{D_S} is generated with the probability $P^{gen}_{|D_S\rangle}\approx
0.85$ -- see Fig.~\ref{LZS_trig}b). The energy gap is larger than in the linear case, because all three hopping parameters $t_{ij}$ are always on. Notice that
for this orientation of the field the qubit state \ket{D_T} is not generated, because the transfer matrix element
$\langle\psi_{in}|\hat{H}|D_T\rangle=\sqrt{3}(t_{23}-t_{12})/2=0$, this state is dark \cite{bulka,luczak}. For the qubit state \ket{D_S} one can found
$\langle\psi_{in}|\hat{H}|D_S\rangle=(t_{23}+t_{12})/2=-1$  for $t_{ij}=-1$.
In Fig.~\ref{LZS_trig}a) one can see also the crossing between the state \ket{D_S} and \ket{D_T} at $g_E=0$
after which the ground state becomes \ket{D_T}. However, there is no mixing between \ket{D_S} and \ket{D_T}, the parameter $\gamma=0$ in the qubit Hamiltonian
(\ref{hamilpaulixyz}), and hence the system is held in $|D_S\rangle$ in whole range of the charge state (1,1,1).
Similarly as for the linear case we can estimate the time for qubit generation. Taking $t_{ij} = 10$ $\mu$eV from the experiment \cite{gaudreau} one gets $\Delta E_{01}=23$
$\mu$eV,  $v= 332.7$ keV/s, and hence the transition time $t^{sw}=0.45$ ns.

The situation is different in the right panel of figure \ref{LZS_trig} when the electric field has opposite orientation. Now at the initial condition for
$g_E=-15$ the dot levels are
$\tilde{\epsilon}_2>\tilde{\epsilon}_1+U_1=\tilde{\epsilon}_3+U_3$ and the ground state is
$|\psi_{in}\rangle=(|D_{102}\rangle-|D_{201}\rangle)/\sqrt{2}$. One can see the initial charge distribution of the function is different than in previous case
--
the dot 2 is empty. The transfer matrix elements are $\langle\psi_{in}|\hat{H}|D_T\rangle=\sqrt{3}(t_{23}+t_{12})/2=-\sqrt{3}$ and
$\langle\psi_{in}|\hat{H}|D_S\rangle=(t_{23}-t_{12})/2=0$ for $t_{ij}=-1$. In this case \ket{D_S} is the dark state and the L--Z transition generates the qubit
state \ket{D_T} (see Fig.\ref{LZS_trig}f). The energy gap is now $E_{01}\approx 3.47$. We take $v= 5.01$ to get the same $\Delta E_{01}^2/v$ rate  with
the generated probability $P^{gen}_{|D_T\rangle}\approx 0.85$. The qubit state is generated in time $t^{sw}= 0.19$ ns.
The period of observed oscillations is larger than for the case $\theta=2\pi/3$ because the gap is larger.
Similarly like in previous case one can see the crossing between \ket{D_T} and \ket{D_S} at $g_E=0$ but now the generated qubit state \ket{D_T} is kept in whole
charge region (1,1,1). We will show latter how to remove this degeneration point and to perform the L--Z passage between the qubit states.

In the middle panel in figure  \ref{LZS_trig} we present an intermediate case for $\theta=5\pi/6$. Now the parameter $\gamma$ is non zero for the whole range of
$g_E$ (except $g_E=0$), which leads to mixing between the doublets \ket{D_S} and \ket{D_T}. The L--Z transition generates the state $|D^-\rangle$ from the
initial state \ket{D_{021}}, with $P^{gen}_{|D^-\rangle}\approx 0.85$ in time $t^{sw} = 0.46$ ns. The occupation probabilities of \ket{D_S} and \ket{D_T} oscillate in phase, see the black
and the red curve in Fig.~\ref{LZS_trig}d). There is no passage to the excited state \ket{D^+} because it
is decoupled with the initial state \ket{D_{021}}.
These results show that generation of the qubit state \ket{D^-} on the Bloch  sphere can be performed by an appropriate orientation of the electric field.

\begin{figure}[h]
\centering
\includegraphics[width=.45\textwidth,clip]{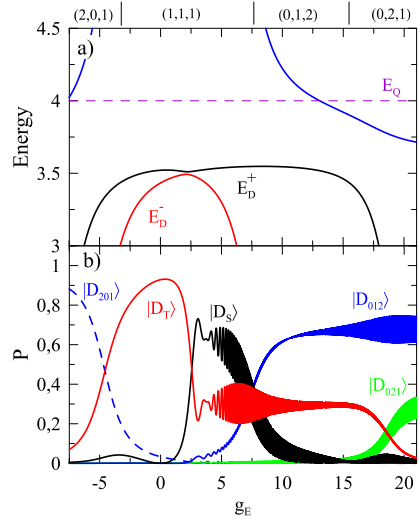}
\caption{Landau--Zener transitions in TQD with the triangular geometry and the second dot being smaller. The electric field is directed toward the dot 1 (the angle  $\theta=0$). Figure is plotted for parameters: $\epsilon_{2}=4$, $U_2=24$, $U_{1,3}=11$,
$\epsilon_{1,3}=0$ and $t_{ij}=-1$. Figure a) presents the energy spectrum for the adiabatic passage, while figure b) the probability of occupations of the doublet states with respect to $g_E=vt$. The speed rate was taken $v=0.01$ to clearly show generation of the qubit states and the Rabi oscillations.}
\label{LZS_trig_asymm}
\end{figure}

Let us now consider the case with one of the dot smaller than two others. We expect that the  symmetry of the system will be broken, the degeneracy of the
doublet states \ket{D^-} and \ket{D^+} will be removed and one can perform the L--Z passage in the center of the charge region (1,1,1). In calculations we took
$\epsilon_2=4$, $\epsilon_{1,3}=0$, $U_2=24$, $U_{1,3}=11$ and $\theta=0$. In Fig.~\ref{LZS_trig_asymm}a)  one can see the adiabatic levels \ket{D^-} and
\ket{D^+} in the region (1,1,1) and an anticrossing point at $g_E\approx 2$. Before this point, for $g_E<2$, the system is in the ground state \ket{D^-} which
is
a superposition \ket{D_T} with a small contribution \ket{D_S}; the qubit is oriented to the south pole on the Bloch sphere. For an adiabatic passage, above
$g_E>2$,  the qubit changes its orientation toward the north pole (\ket{D_S} becomes dominating). The results presented in Fig.~\ref{LZS_trig_asymm}b) were
calculated for the L--Z transition with the speed rate $v=0.01$.
One can see generations of the excited state \ket{D^+} and the Rabi oscillations (oscillations between \ket{D_T} and \ket{D_S}. After anti-crossing the fidelity for \ket{D_S} is $F\approx 0.93$.
This effect can be used to dynamic control the qubit states and to perform quantum operations \cite{laird10}.
Plots in Fig.~\ref{LZS_trig_asymm}b) show also two other L--Z transitions: at $g_E \approx 7.5$ and $g_E \approx 18$ where the states \ket{D_{012}} and
\ket{D_{021}} are generated, respectively. Notice that \ket{D_T} is kept till the transition at $g_E\approx 18$.

\section{Conclusions}

Summarizing, we considered the system of
three coherently coupled quantum dots (TQD) in the linear and the triangular geometry, for which quantum dot levels were
controlled by the local gate potentials.  Applying electrical pulses the Landau--Zener transitions were generated between different charge configurations
$(N_1,N_2,N_3)$ of the system. The
spin qubit states were encoded in the doublet subspace \ket{D_S} and \ket{D_T} in the charge region (1,1,1), and they were generated from the initial charge
states with
double dot occupancy (e.g. from $(2,1,0)$).

Our research on the linear TQD expands the investigations presented in \cite{aers,laird10,gaudreau}. We showed that in this case the state \ket{D_T} is
preferred, the   generated qubit is restricted to the south part of the Bloch sphere only. For Si-based quantum dots \cite{zwanenburg,culcer} the estimated sweep time $t^{sw}$ for the generation of the qubit is of the order of a few nanoseconds which is much shorter than both the decoherence and relaxation time. We also considered the system with one of the dots smaller than two others. The qubit states can not be well generated, then the fidelity is rather low and the final state always contains a large
superposition of non-qubit states for any speed rate.

In the triangular geometry  both the qubit states are equivalent and can be easily generated by the L--Z transitions for the mirror symmetries.
The symmetry was changed by a rotation of the effective electric field (controlled by the local gate potentials applied to the quantum dots). If the electric field is oriented toward the second quantum dot, then the L--Z
transition generates the qubit state \ket{D_S} from the initial state $|\psi_{in}\rangle=(|D_{021}\rangle+|D_{120}\rangle)/\sqrt{2}$.
For the opposite direction of the electric field, the second qubit state \ket{D_T} is obtained from the initial state
$|\psi_{in}\rangle=(|D_{102}\rangle-|D_{201}\rangle)/\sqrt{2}$. The L-Z transition induces characteristic oscillations between the initial and the qubit state
with the amplitude related to the speed rate $v$. For small $v$ these oscillations can be suppressed and the qubit is generated with a higher fidelity.
The energy gaps are larger than in the linear molecule which allows faster generation of the qubit states (with $t^{sw}\approx$ 0.1 ns). We also showed that one can
get any qubit state
on the Bloch sphere by a proper orientation of the electric field.
The interested case is the triangular TQD with one of the dots smaller for which an additional anticrossing point between the qubit states occurs. We showed that
the L--Z
passage through this point generates the Rabi oscillations. This effect can be used to perform one-qubit operations. Applying a proper sequence of electrical pulses one can perform the Pauli X and Z gate which give full unitary control
of the qubit rotation on the Bloch sphere \cite{luczak}.

Our calculations showed that the dynamics of the qubit should be observable in an experiment on Si-based quantum dots even in the presence of the decoherence processes for reasonable transition speeds.

The triangular TQD system can be used for construction of a multi-qubit quantum register with fast quantum logical operations. Changing the symmetry of each qubit one can easily encode a desire initial state in the register. For the linear TQD the encoding operation is more complex, because one should perform an additional single-qubit gate in order to rotate the qubit states, which significantly increases the operation time.

\begin{acknowledgements}
This work was supported by the National Science Centre under the contract  DEC-2012/05/B/ST3/03208.
\end{acknowledgements}


\begin{thebibliography}{99}

\bibitem{vrijen} Vrijen, R., Yablonovitch, E., Wang, K., Jiang, H. W., Balandin, A., Roychowdhury,  V., Mor, T., DiVincenzo, D.: Electron-spin-resonance transistors for quantum computing in silicon-germanium heterostructures, Phys. Rev. A {\bf 62}, 012306 (2000).

\bibitem{engel} Engel, H.-A., Loss, D.: Single-spin dynamics and decoherence in a quantum dot via charge transport, Phys. Rev. B {\bf 65}, 195321 (2002).

\bibitem{nowack} Nowack, K. C., Koppens,  F. H. L., Nazarov,  Yu. V., Vandersypen, L. M. K.: Coherent Control of a Single Electron Spin with Electric Fields, Science {\bf 318}, 1430 (2007).

\bibitem{veldhorst} Veldhorst, M., Yang, C. H., Hwang, J. C. C., Huang, W., Dehollain, J. P., Muhonen, J. T., Simmons, S., Laucht, A., Hudson, F. E., Itoh, K. M., Morello, A., Dzurak, A. S.: A two-qubit logic gate in silicon, Nature {\bf 526}, 410 (2015).

\bibitem{loss} Loss, D., DiVincenzo, D. P.: Quantum computation with quantum dots, Phys. Rev. A {\bf 57}, 120 (1998).

\bibitem{petta2005} Petta,  J. R., Johnson,  A. C., Taylor,  J. M., Laird,  E. A., Yacoby,  A., Lukin,  M. D., Marcus,  C. M., Hanson,  M. P., Gossard,  A. C.: Coherent Manipulation of Coupled Electron Spins in Semiconductor Quantum Dots, Science {\bf309},
    2180 (2005).

\bibitem{bluhm} Bluhm, H., Foletti, S., Neder, I., Rudner, M., Mahalu, D., Umansky, V., Yacoby, A.: Dephasing time of GaAs electron-spin qubits
coupled to a nuclear bath exceeding 200 $\mu$s,  Nature Phys. {\bf 7}, 109 (2010).

\bibitem{sarkka} S\"{a}rkk\"{a}, J., Harju, A.: Spin dynamics at the singlet–triplet crossings in a double quantum dot, New J. Phys. {\bf 13}, 043010 (2011).

\bibitem{granger} Granger, G., Aers, G. C., Studenikin, S. A., Kam, A., Zawadzki, P., Wasilewski, Z. R., Sachrajda, A. S.: Visibility study of $S - T_+$ Landau-Zener-Stückelberg oscillations without applied initialization, Phys. Rev. B {\bf 91}, 115309 (2015);
                  Studenikin, S. A., Aers,  G. C., Granger,  G., Gaudreau,  L., Kam, A., Zawadzki, P., Wasilewski, Z. R., Sachrajda, A. S.: Quantum Interference between Three Two-Spin States in a Double Quantum Dot, Phys. Rev. Lett. {\bf 108}, 226802 (2012).

\bibitem{dial} Dial, O. E., Shulman, M. D., Harvey, S. P., Bluhm, H., Umansky, V., Yacoby, A.: Charge Noise Spectroscopy Using Coherent Exchange Oscillations in a Singlet-Triplet Qubit, Phys. Rev. Lett. {\bf 110}, 146804 (2013).

\bibitem{wu} Wu, X., Ward, D. R., Prance, J. R., Kim, D., Gamble, J. K., Mohr, T. R., Shi, Z., Savage, D. E., Lagally, M. G., Friesen, M., Coppersmith, S. N., Eriksson, M. A.: Two-axis control of a singlet–triplet qubit with an integrated micromagnet, Proc. Natl. Acad. Sci. USA {\bf 111}, 11938 (2014).

\bibitem{barthel} Barthel, C., Reilly, D. J., Marcus, C. M., Hanson, M. P., Gossard, A. C.: Rapid Single-Shot Measurement of a Singlet-Triplet Qubit, Phys. Rev. Lett. {\bf 103}, 160503 (2009).

\bibitem{maune} Maune, B. M., Borselli, M. G., Huang, B., Ladd, T. D., Deelman, P. W., Holabird, K. S., Kiselev, A. A., Alvarado-Rodriguez, I., Ross, R. S., Schmitz, A. E., Sokolich, M., Watson, C. A., Gyure, M. F., Hunter, A. T.: Coherent singlet-triplet oscillations in a silicon-based double quantum dot, Nature {\bf 481}, 344 (2012).

\bibitem{shevchenko} Landau, L.: Zur Theorie der Energieubertragung. II, Phys. Z. Sowjetunion {\bf 2}, 46 (1932);
Zener, C.: Non-Adiabatic Crossing of Energy Levels, Proc. R. Soc. A {\bf 137} 696 (1932);
    Shevchenko, S. N., Ashhab, S., Nori, F.: Landau–Zener–Stückelberg interferometry, Physics Reports {\bf 492} 1-30 (2010).

\bibitem{hanson} Hanson, R., Kouwenhoven, L. P., Petta, J. R., Tarucha, S., Vandersypen, L. M.: Spins in few-electron quantum dots, Rev. Mod. Phys. {\bf 79}, 1217 (2007).

\bibitem{bason} Bason, M. G., Viteau, M., Malossi, N., Huillery, P., Arimondo, E., Ciampini, D., Fazio, R., Giovannetti, V., Mannella, R., Morsch, O.: High-fidelity quantum driving, Nature Phys. {\bf 8}, 147 (2012).

\bibitem{hicke} Hicke, C., Santos, L. F., Dykman, M. I.: Fault-tolerant Landau-Zener quantum gates, Phys. Rev. A, {\bf 73} 012342 (2006).

\bibitem{nichol} Nichol, J. M., Harvey, S. P., Shulman, M. D., Pal,	A., Umansky, V., Rashba, E. I., Halperin, B. I., Yacoby, A.: Quenching of dynamic nuclear polarization by spin–orbit coupling in GaAs quantum dots,  Nature Comm. {\bf 6}, 7682 (2015).

\bibitem{vincenzo} DiVincenzo, D. P., Bacon, D., Kempe, J., Burkard, G., Whaley, K. B.: Universal quantum computation with the exchange interaction, Nature (London) {\bf408}, 339 (2000).

\bibitem{dfs} Lidar, D. A., Chuang, I. L., Whaley, K. B.: Decoherence-Free Subspaces for Quantum Computation, Phys. Rev. Lett. {\bf 81}, 2594 (1998);
Bacon, D., Kempe, J., Lidar, D. A., Whaley, K. B.: Universal Fault-Tolerant Quantum Computation on Decoherence-Free Subspaces, Phys. Rev. Lett. {\bf 85}, 1758 (2000).


\bibitem{fei} Fei, J., Hung, J.-T., Koh, T. S., Shim, Y.-P., Coppersmith, S. N., Hu, X., Friesen, M.: Characterizing gate operations near the sweet spot of an exchange-only qubit, Phys. Rev. B {\bf 91}, 205434 (2015).

\bibitem{pal} Pal, A., Rashba, E. I., Halperin, B. I.: Driven Nonlinear Dynamics of Two Coupled Exchange-Only Qubits, Phys. Rev. X {\bf 4}, 011012 (2014).

\bibitem{busl} Busl, M., S\'{a}nchez, R., Platero, G.: Control of spin blockade by ac magnetic fields in triple quantum dots, Phys. Rev. B {\bf 81}, 121306(R) (2010).

\bibitem{busl2013} Busl, M., Granger, G., Gaudreau, L., S\'{a}nchez, R., Kam, A., Pioro-Ladriere, M., Studenikin, S. A., Zawadzki, P., Wasilewski, Z. R., Sachrajda, A. S., Platero, G.: Bipolar spin blockade and coherent state superpositions in a triple quantum dot, Nature Nanotech. {\bf 8}, 261 (2013).

\bibitem{aers} Aers, G. C., Studenikin, S. A., Granger, G., Kam, A., Zawadzki, P., Wasilewski, Z. R., Sachrajda, A. S.: Coherent exchange and double beam splitter oscillations in a triple quantum dot, Phys. Rev. B {\bf 86}, 045316 (2012).

\bibitem{laird10} Laird, E. A., Taylor, J. M., DiVincenzo, D. P., Marcus, C. M., Hanson, M. P., Gossard, A. C.: Coherent spin manipulation in an exchange-only qubit, Phys. Rev. B {\bf82} 075403 (2010).

\bibitem{gaudreau} Gaudreau, L., Granger, G., Kam, A., Aers, G. C., Studenikin, S. A., Zawadzki, P., Pioro-Ladriere, M., Wasilewski, Z. R., Sachrajda, A. S.: Coherent control of three-spin states in a triple quantum dot, Nature Phys. {\bf 8}, 54 (2012).

\bibitem{taylor} Taylor, J. M., Srinivasa, V., Medford, J.: Electrically Protected Resonant Exchange Qubits in Triple Quantum Dots, Phys. Rev. Lett. {\bf 111}, 050502 (2013).

\bibitem{medford} Medford, J., Beil, J., Taylor, J.M., Rashba, E.I., Lu, H., Gossard, A.C., Marcus, C.M.: Quantum-Dot-Based Resonant Exchange Qubit, Phys. Rev. Lett. {\bf 111}, 050501 (2013).

\bibitem{shi} Shi, Z., Simmons, C. B., Prance, J. R., Gamble, J. K., Koh, T. S., Shim, Y.-P., Hu, X., Savage, D. E., Lagally, M. G., Eriksson, M. A., Friesen, M., Coppersmith, S. N.: Fast Hybrid Silicon Double-Quantum-Dot Qubit, Phys. Rev. Lett. {\bf 108}, 140503 (2012).

\bibitem{hawrylak} Hawrylak, P., Korusinski, M.: Voltage-controlled coded qubit based on electron spin, Solid State Commun. {\bf 136}, 508 (2005).

\bibitem{bulka} Bu{\l}ka, B. R., Kostyrko, T., {\L}uczak, J.: Linear and nonlinear Stark effect in a triangular molecule,  Phys. Rev. B {\bf83}, 035301 (2011).

\bibitem{luczak} {\L}uczak, J., Bu{\l}ka, B. R.: Readout and dynamics of a qubit built on three quantum dots, Phys. Rev. B {\bf 90} 165427 (2014).

\bibitem{rogge} Rogge, M. C., Haug, R. J.: Two-path transport measurements on a triple quantum dot, Phys. Rev. B {\bf 77}, 193306 (2008);
Rogge, M. C., Haug, R. J.: Noninvasive detection of molecular bonds in quantum dots, Phys. Rev. B {\bf 78}, 153310 (2008);
Rogge, M. C., Haug, R. J.: The three dimensionality of triple quantum dot stability diagrams, New J. Phys. {\bf 11}, 113037 (2009).

\bibitem{seo} Seo, M., Choi, H. K., Lee, S.-Y., Kim, N., Chung, Y., Sim, H.-S., Umansky, V., Mahalu, D.: Charge Frustration in a Triangular Triple Quantum Dot, Phys. Rev. Lett. {\bf 110}, 046803 (2013).

\bibitem{amaha09} Amaha, S., Hatano, T., Kubo, T., Teraoka, S., Tokura, Y., Tarucha, S., Austing, D. G.: Stability diagrams of laterally coupled triple vertical quantum dots in triangular arrangement, Appl. Phys. Lett. {\bf 94}, 092103 (2009).


\bibitem{molmag} Trif, M., Troiani, F., Stepanenko, D., Loss, D.: Spin-Electric Coupling in Molecular Magnets, Phys. Rev. Lett. {\bf 101}, 217201 (2008);
B. Tsukerblat, Inorg. Chim. Acta {\bf 361}, 3746 (2008).

\bibitem{zwanenburg} Zwanenburg, F. A., Dzurak, A. S., Morello, A., Simmons, M. Y., Hollenberg, L.d C. L., Klimeck, G., Rogge, S.,. Coppersmith, S. N, Eriksson, M. A.: Silicon quantum electronics, Rev. Mod. Phys. {\bf 85}, 961 (2013).

\bibitem{simmons} Simmons, C. B., Prance, J. R., Van Bael, B. J., Teck Seng Koh, Zhan Shi, Savage, D. E., Lagally, M. G., Joynt, R., Friesen, M., Coppersmith, S. N., Eriksson, M. A.: Tunable Spin Loading and $T_1$ of a Silicon Spin Qubit Measured by Single-Shot Readout, Phys. Rev. Lett. {\bf 106}, 156804 (2011).

\bibitem{kostyrko} Kostyrko, T., Bu{\l}ka, B. R.: Canonical perturbation theory for inhomogeneous systems of interacting fermions, Phys. Rev. B {\bf 84}, 035123 (2011).

\bibitem{shi2013} Shi, Z., Simmons, C. B., Ward, D. R., Prance, J. R., Mohr, R. T., Koh, T. S., Gamble, J. K., Wu, X., Savage, D. E., Lagally, M. G., Friesen, M., Coppersmith, S. N., Eriksson, M. A.: Coherent quantum oscillations and echo measurements of a Si charge qubit, Phys. Rev. B {\bf 88}, 075416 (2013).

\bibitem{culcer} Culcer, D., Hu, X., Das Sarma, S.: Dephasing of Si spin qubits due to charge noise, Appl. Phys. Lett. {\bf 95}, 073102 (2009).

\bibitem{hauke} Hauke, P., Cucchietti, F. M., Tagliacozzo, L., Deutsch, I., Lewenstein, M.: Can one trust quantum simulators?, Rep. Prog. Phys. {\bf 75}, 082401 (2012).

\bibitem{kayanuma} Kayanuma, Y., Fukuchi, S.: On the probability of non-adiabatic transitions in multiple level
crossings, J. Phys. B: At. Mol. Phys. {\bf 18}, 4089 (1985); Shytov, A. V.: Landau-Zener transitions in a multilevel system: An exact result, Phys. Rev. A {\bf 70}, 052708(2004).


\end{thebibliography}
\end{document}